# Evidence of Two-Dimensional Quantum Critical Behavior in the Superfluid Density of Deeply Underdoped $Bi_2Sr_2CaCu_2O_{8+x}$

**Jie Yong,[1,*] M. J. Hinton[1], A. McCray,[1] M. Randeria[1], M. Naamneh,[2] A. Kanigel,[2] and T. R. Lemberger[1]**

[1]Department of Physics, The Ohio State University, Columbus, OH, USA

[2]Department of Physics, Technion - Israel Institute of Technology, Haifa 32000, Israel

[*]E-mail: yong.13@osu.edu

Evidence of two-dimensional (2D) quantum critical fluctuations is observed in the superfluid density $n_s(T) \propto \lambda^{-2}(T)$ of deeply underdoped $Bi_2Sr_2CaCu_2O_{8+x}$ (Bi-2212) films, indicating that quantum fluctuations play a dominant role in underdoped cuprates in general. 2D fluctuations are expressed by the linear scaling, $T_c \propto n_s(0)$. 2D scaling in Bi-2212 contrasts with 3D scaling seen in the much less anisotropic $YBa_2Cu_3O_{7-\delta}$. Quantum critical fluctuations could also account for the absence of thermal critical behavior in $\lambda^{-2}(T)$ of strongly underdoped Bi-2212 samples, $T_c < 48$ K.

The underdoped side of cuprates, due to its proximity to antiferromagnetism and the mysterious pseudogap, is the least-understood part of the cuprate phase diagram and has been the focus of much high temperature superconductivity research in recent years. It is interesting to ask why and how superconductivity disappears when cuprates are severely underdoped. Superfluid density measurements, which probe the temperature dependence of superconducting carrier density and thus its vulnerability to thermal and quantum fluctuations, can provide insight to this question. Measurements on severely underdoped $YBa_2Cu_3O_{7-\delta}$ (YBCO) crystals [1, 2] and films [3] show that superconductivity in this compound disappears at a 3D quantum critical point. First, there is a sublinear scaling relationship, $T_c \propto n_s(0)^\alpha$ with $\alpha \sim ½$, indicating strong 3D quantum phase fluctuations (with a dynamical critical exponent, $z_Q \approx 1$) [4, 5]. In addition, with underdoping, $n_s(T)$ loses its downward curvature near $T_c$, which appears to be a loss of thermal critical behavior. This loss seems counter intuitive because underdoping tends to make cuprates even more anisotropic, which favors thermal critical behavior. Interestingly, 2D Ca-doped YBCO films, which, appropriately, exhibit 2D quantum critical scaling, also exhibit strong 2D thermal critical behavior [6].

There are two strong motivations for studying the compound $Bi_2Sr_2CaCu_2O_{8+x}$ (Bi-2212). First, this compound is much more anisotropic than YBCO, so qualitatively different results could emerge. The resistivity anisotropy for Bi-2212 is in the order of $10^5$ [7, 8] while it is only about $10^2$ for YBCO [9]. For example, will quantum critical fluctuations be present, and if so, will they be 2D amid extreme anisotropy or will they be 3D, like YBCO? Second, Bi-2212 is the workhorse compound for studies of the superconducting gap function via angle-resolved photoelectron spectroscopy (ARPES) and scanning tunneling microscopies (STM). Those surface-sensitive measurements heavily focus on Bi-2212 because it can be cleaved easily between the very weakly bonded Bi-O layers. In order to correlate the evolution of quantum and thermal critical fluctuations with the evolution of the superconducting gap function, we best study Bi-2212.

Our samples are Bi-2212 films grown by pulsed laser deposition (PLD) onto MgO substrates (at OSU) and by sputtering onto $LaAlO_3$ substrates (at Technion). Targets are stoichiometric $Bi_2Sr_2CaCu_2O_{8+x}$. Hole doping is tuned by oxygen pressure during deposition and postannealing. All the films are epitaxial and phase pure as indicated by X-ray diffraction measurements. Full Width at Half Maximum (FWHM) values of rocking curves are always 0.3 to 0.4 degrees. Severely underdoped films tend to have slightly larger out-of-plane lattice constants than moderately underdoped films, a fact which agrees with YBCO data [10]. All the films are about 100 nm thick. Transition temperature varies from $T_{c,max}$~80K for near optimal doped films to $T_{c, min}$ ~ 6K for severely underdoped ones in steps of about 5K. The good control of doping level is unprecedented and allow let us observe precisely how superfluid density evolves with doping until superconductivity is destroyed. Data from different types of samples overlap pleasantly, as discussed below.

Superfluid densities are measured by a two-coil mutual inductance apparatus [11, 12]. The film is sandwiched between two coils, and the mutual inductance between these two coils is measured at a frequency $\omega/2\pi$ = 50 kHz. The measurement actually determines the sheet conductivity, $Y \equiv (\sigma_1 + i\sigma_2)d$, with $d$ being the superconducting film thickness and $\sigma$ being the conductivity. Given a measured film thickness, $\sigma$ is calculated as: $\sigma = Y/d$. The imaginary part, $\sigma_2$, yields the superfluid density through: $\omega\sigma_2 \equiv n_s e^2/m$, which is proportional to the inverse penetration depth squared: $\lambda^{-2}(T) \equiv \mu_0\omega\sigma_2(T)$, where $\mu_0$ is the permeability of vacuum. As is customary, we refer to $\lambda^{-2}$ as the superfluid density. The dissipative part of the conductivity, $\sigma_1(T)$, has a peak near $T_c$, whose width provides an upper limit on the spatial inhomogeneity of $T_c$ over the ~10 mm$^2$ area probed by the measurement. Data are taken continuously as the sample slowly warms up so as to yield the hard-to-measure absolute value of $\lambda$ and its T-dependence. This two-coil technique is especially powerful [13] for studies of thermal critical behavior near $T_c$ and quantum critical scaling because $\lambda^{-2}(0)$ can be measured.

The great advantage of studying Bi-2212 films, as opposed to bulk, is that we are able to underdope them by reducing oxygen concentration to much lower $T_c$'s than is possible in bulk. The substrate probably provides mechanical stability. As discussed below, our films agree with bulk samples at dopings where data on powder samples exist, suggesting that we are measuring the properties of Bi-2212 in general, and that our results are insensitive to whatever flaws exist in films. To back up this notion, PLD and sputtered films with similar $T_c$'s have similar superfluid densities both in magnitude and T-dependence, even though these films likely have different types of structural and chemical flaws, given the differences in deposition process and in substrate material.

Figure 1 shows $\lambda^{-2}(T)$ of many Bi-2212 films. Red/pink curves are sputtered films and blue/green curves are PLD films. The narrow widths of the peaks in $\sigma_1(T)$ indicate all the films have decently sharp transitions, with transition widths of about 2 K for most samples. Sputtered and PLD films agree well with each other, as seen from pairs of films with $T_c$'s near 53 K and 37 K. Also, note the reproducibility of two PLD films that happen to have the same $T_c$ of 28 K. When we compare our data with the only other available superfluid density data, which are on underdoped Bi-2212 powders, we find that scaling of $T_c$ with $n_s(0)$ is similar for our films and Bi-2212 powders at moderate underdoping [14], which shows that $n_s(0)$ decreases much faster than $T_c$ as films get underdoped. However, powder data do not show sharp downturns near $T_c$ as we observed in Bi-2212 films. Also notice that powders and films give the same superfluid density, $n_s(0) \propto \lambda^{-2}(0) \sim 1 \mu m^{-2}$, when $T_c$ is around 40K. This is remarkable given the fact that these are two totally different measurements on very different samples (films vs. powders).

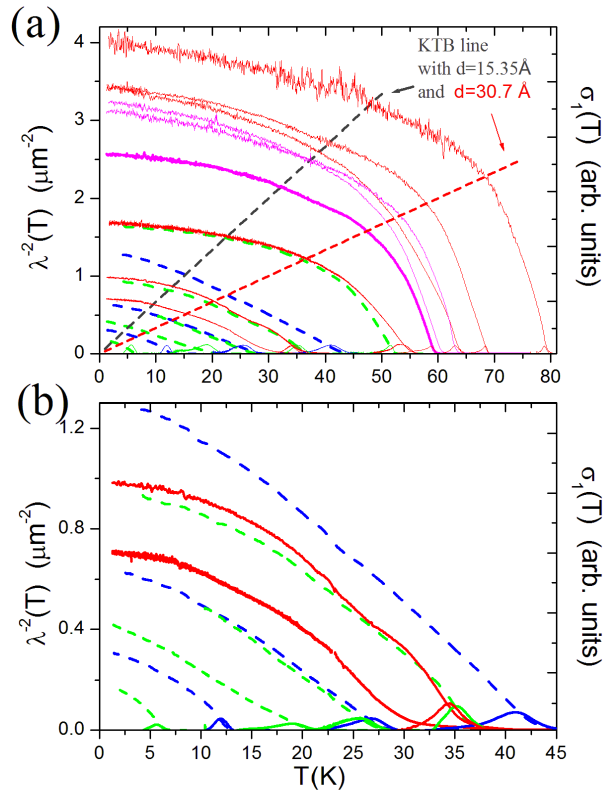

FIG. 1. (Color online): T-dependence of superfluid density for underdoped Bi-2212 films (PLD films in dashed green/blue; sputtered films in solid red/pink). (a) full range of dopings. Intersection of upper (lower) dashed line with $\lambda^{-2}(T)$ is where Kosterlitz-Thouless theory predicts a downturn in $\lambda^{-2}$ due to vortex-antivortex unbinding for d=15.35 Å (1 $CuO_2$ bilayer) [d=30.7 Å (2 $CuO_2$ bilayers)]. (b) severely underdoped films with $T_c$ < 45 K. The widths of peaks in $\sigma_1$ near $T_c$ set an upper limit on the spatial inhomogeneity of $T_c$.

Consistent features for moderately underdoped films, $T_c$ > 50 K, are the weak linear T-dependence at low-T and sharp downturn near $T_c$. Given the extremely anisotropic nature of Bi-2212, we associate this downturn with quasi-2D thermal phase fluctuations developing in individual $CuO_2$ bilayers. For 2D superconductors, Kosterlitz-Thouless (KT) theory [15] predicts a downturn where $k_B T$, ($k_B$ = Boltzmann's constant) is about equal to the energy required to create a vortex-antivortex pair, i.e., the transition occurs at the temperature where: $\lambda^{-2}(T) = (2\mu_0/\pi d \hbar R_Q) k_B T$. The quantum resistance $R_Q \equiv \hbar/4e^2$ ~1 kΩ, $d$ is the effective 2D thickness. The two dashed lines in Fig. 1 represent the right hand side of this equation assuming that $d$ = 15.35 Å (1 $CuO_2$ bilayer) and $d$ = 30.7 Å (2 $CuO_2$ bilayers). Intersection of the latter line with the observed $\lambda^{-2}(T)$ is consistent with what is seen in 2 unit cell thick YBCO films [6] and conventional 2D superconducting films [16]. Thus, we conclude that the extreme anisotropy of Bi-2212 brings the effective 2D layer thickness down to 2 bilayers as regards classical thermal fluctuations.

The observation of 2-D thermal critical behavior in moderately underdoped Bi-2212 is consistent with early results that show a frequency-dependent superfluid density above $T_c$ [17] and a nonlinear magnetization above $T_c$ which is interpreted by BKT physics [18]. It is also interesting to note that this 2-D thermal critical behavior is not observed in other cuprates, like in $YBa_2Cu_3O_{7-x}$ [19, 20] or $La_{2-x}Sr_xCuO_4$ [21], where a 3-D XY or mean field like behavior is observed. This qualitative difference is probably coming from the huge difference in anisotropy in Bi-2212 compared to other cuprate materials.

It is odd that the sharp KT-like downturn in $\lambda^{-2}(T)$ diminishes with severe underdoping. Nominally, a reduction in superfluid density should enhance critical phase fluctuations. Also odd is the abruptness of the downturn's disappearance, which occurs when $T_c$ drops below about 48K. As seen below, this is the same place where scaling of $T_c$ with $\lambda^{-2}(0)$ becomes linear. As for the former, Franz and Iyengar [5] have explained that the thermal critical regime can narrow dramatically for samples near a quantum critical point (QCP), thereby accounting for the disappearance of downward curvature near $T_c$. If the downturn in $\lambda^{-2}(T)$ for moderate underdoping is due to critical thermal phase fluctuations, and its disappearance is due to quantum fluctuations near a QCP, it is natural to suppose that quantum and thermal phase fluctuations conspire to dominate the T-dependence of $\lambda^{-2}(T)$ for severely underdoped films. The following analysis shows that this notion is reasonable, with some caveats.

Classical thermal phase fluctuations are expected to suppress superfluid density by a factor of $(1 - k_B T/4J)$ in square arrays of superconducting grains coupled by Josephson energy J, as long as the suppression is small [23]. For continuous films, the effective coupling energy J is proportional to the sheet superfluid density: $J(T) \equiv \hbar^2 n_s^{2D}/4m$, where sheet superfluid density $n_s^{2D}(T) = n_s(T)*d$ and $n_s(T)/m = \lambda^{-2}(T)/\mu_0 e^2 = 4R_Q \lambda^{-2}(T)/\hbar \mu_0$, therefore $J(T) = \hbar R_Q d\lambda^{-2}(T)/\mu_0$, where $d$ is the effective 2D thickness. Thus, the low-T behavior would be:

$$\lambda^{-2}(T) \approx \lambda^{-2}(0)[1 - k_B T/4J(0)] = \lambda^{-2}(0) - \mu_0 k_B T/4\hbar R_Q d.$$

Thus, the low-T slope of $\lambda^{-2}(T)$ does not depend on $\lambda^{-2}(0)$ in this model. Quantum effects would change linear to quadratic behavior below some cut-off temperature [24], but we set that physics aside. We see in the measured $\lambda^{-2}(T)$ for films with $T_c < 48$ K that the slopes of the quasilinear data are, indeed, roughly independent of $\lambda^{-2}(0)$. Quantitatively, the slope should be -0.013 $\mu m^{-2}$/K (or -0.0065 $\mu m^{-2}$/K), assuming $d = 30.5$ Å (or 15.25 Å), which is in the range of the measured slope (~0.008$\mu m^{-2}$/K). Of course, we have used a low-T expression to analyze data up to $T_c$. This might indicate another energy scale which is much higher than $T_c$. Further theoretical work is needed to determine whether quantum critical fluctuations would repair this discrepancy.

Although anomalous, the quasi-linear T-dependence of $\lambda^{-2}$ (from about $T_c/3$ to $T_c$) at severe underdoping is also seen in severely underdoped YBCO crystals [1, 2] and films [3]. At the lowest temperatures, all of these superfluid densities tend to level off and this can be understood in the context of d-wave pairing symmetry with significant impurity scattering, which is similar to what has been observed in YBCO crystals. [1] Apparently this linear behavior is robust against anisotropy and disorder. It may be regarded as a universal phenomenon for severely underdoped cuprates, and it therefore provides important guidance to theory. For example, Igor Herbut started with the field theory of a fluctuating d-wave superconductor and interpreted this quasi-linear T-dependence of superfluid density in the context of a strongly anisotropic weakly interacting Bose gas [22].

Having argued for the presence of strong quantum critical fluctuations on the basis of absence of thermal critical behavior, we turn to a second key indicator, namely, power-law scaling of $T_c$ with $\lambda^{-2}(0)$. This scaling, if present, is sensitive to dimensionality. If quantum fluctuations are 2D, then theory of quantum critical scaling [4] requires that $T_c \sim [\lambda^{-2}(0)]^\alpha$ where $\alpha = 1$, regardless of details.

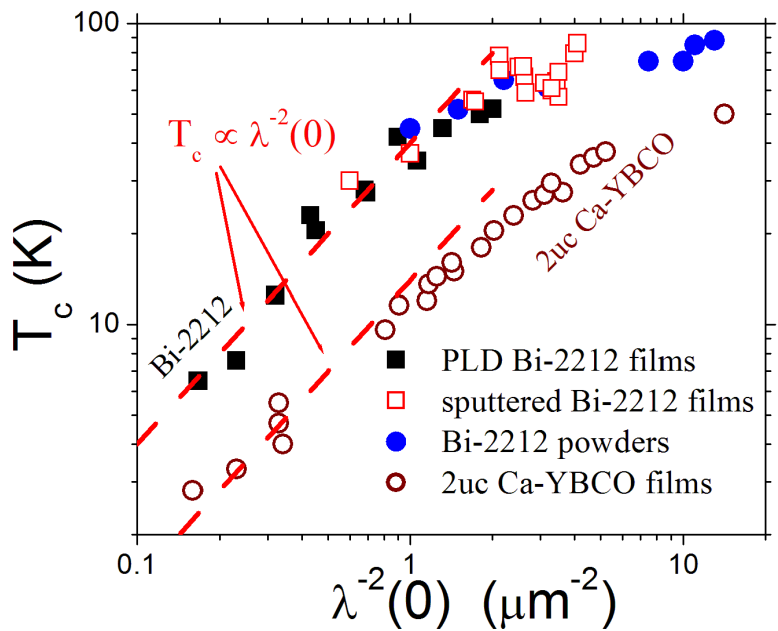

FIG. 2. (Color online): Scaling between $T_c$ and $\lambda^{-2}(0)$ for underdoped Bi-2212 films and powders (PLD films – filled blue squares; sputtered films – open red squares; powders [14] – filled black circles). Data for 2-unit-cell thick Ca-YBCO films (open red circles) are plotted in comparison. Two dashed lines show linear relationships between $T_c$ and $\lambda^{-2}(0)$.

Fig. 2 is a log-log plot of $T_c$ vs. $\lambda^{-2}(0)$ for Bi-2212 films and powders [14]. Data on two-unit-cell thick YBCO films show the behavior of a truly 2D cuprate. Note that Bi-2212 films agree with Bi-2212 powders down to the lowest dopings achieved in powders, $T_c \approx 40$ K and $\lambda^{-2}(0) \approx 1$ μm$^{-2}$. Also, sputtered and PLD films agree with each other, indicating insensitivity to microscopic details. Overall scaling of $T_c$ vs. $\lambda^{-2}(0)$ is qualitatively the same in Bi-2212 and in 2uc YBCO. In particular, scaling is linear at the lowest dopings, indicating that quantum fluctuations are 2D in Bi-2212 probably due to its extremely high anisotropy. In YBCO scaling is square-root, indicating 3D quantum fluctuations [1-3].

The quasi-2D nature of Bi-2212 is apparent in other measurements, e.g., $T_c$ is unaffected by intercalation of various materials into BiO bilayers [25], reducing coupling between adjacent $CuO_2$ bilayers. Setting aside theory, the semi-quantitative similarity between Bi-2212 and 2-unit-cell thick YBCO films (Fig. 2) indicates similar fluctuation physics. The main qualitative difference is that Kosterlitz-Thouless physics is seen in the T-dependence of $\lambda^{-2}(T)$ for severely underdoped 2D YBCO films but not for Bi-2212. Apparently the very weak interlayer coupling in Bi-2212 is enough to permit quantum fluctuations to narrow the critical thermal region beyond visibility.

To put our work into context, we compare this work with previous work on severely underdoped YBCO (thick films or crystals) and ultrathin 2 unit cell YBCO films. It seems that superconductivity always disappears at a quantum critical point when cuprates are severely underdoped, although there are dimensionality differences due to different anisotropy level. For extremely anisotropic Bi-2212 and ultrathin YBCO films, it is a 2-D QCP, while for less anisotropic YBCO, it is a 3-D QCP. Despite different thermal critical behaviors near optimal doping due to anisotropy differences [19], both Bi-2212 (films) and YBCO (films and crystals) lose this thermal critical behavior because of the shrinking of thermal critical region near the quantum critical point. As a result, both of their superfluid densities become quasi-linear all the way to $T_c$. This can be regarded as a universal behavior for severely underdoped cuprates, regardless of differences in anisotropy and disorder. However, for ultrathin YBCO films which are 2-D by construction, thermal critical behavior, a sharp downturn near $T_c$, persists to the lowest doping level. This is inconsistent with the picture that thermal critical behavior shrinks near a QCP. New physics is needed to resolve this discrepancy. Our work shows how thermal and quantum fluctuations, in the context of different anisotropy and disorder level, conspire to destroy superconductivity in severely underdoped cuprates near a quantum critical point.

In summary, by studying superfluid density $n_s(T)$ of underdoped $Bi_2Sr_2CaCu_2O_{8+x}$ with unprecedented control on doping level, we observe strong 2-D quantum fluctuations in strongly underdoped films, evident by linear scaling between $T_c$ and $n_s(0)$. This quantum critical scaling is accompanied by the disappearance of thermal critical behavior near $T_c$. We conclude these 2-D quantum fluctuations are responsible for the disappearance of superconductivity in deeply underdoped Bi-2212.

Work at OSU was supported by DOE-Basic Energy Sciences through grant FG02-08ER46533 (JY, MJH, AM, TRL), by the OSU Dept. of Physics (AM), by DE-SC0005035 (MR) and by the Center of for Emergent Materials, an NSF MRSEC grant No. DMR-0820414. Work at Technion was supported by the Israeli Science Foundation.